\documentstyle[epsf]{l-aa}     
\def\apj{ApJ\ }
  
\def\apjs{ApJS\ }
   
\def\mn{MNRAS }   
\def\etal{{\it et al.~}} 
\def\ha{hereafter~} 
\def\ee{EETDA}
\def\Mpc{$h^{-1}$~Mpc}
\def\msn{\par\nobreak\noindent}

\begin{document}
 
   \thesaurus{12 (12.03.3; 12.12.1) }      
\title{ The Supercluster--Void Network. }
\subtitle{ IV. The Shape and Orientation of Superclusters}
\author{ Jaak Jaaniste\inst{1}, Erik Tago\inst{1}, Maret
Einasto\inst{1}, Jaan Einasto\inst{1},  Heinz Andernach\inst{2}
and Volker M\"uller\inst{3}} 

\institute{ $^{1}$ Tartu Observatory, EE-2444 T\~oravere, Estonia\\
            $^{2}$ Depto. de Astronom\'\i a, IFUG, Apdo. Postal 144, 
            36000 Guanajuato, Mexico\\
          $^{3}$ Astrophysical Institute Potsdam, An der Sternwarte 16,
            D-14482 Potsdam, Germany }
\offprints{J. Jaaniste }
\date{Received ... 1997 / Accepted ... 1997} 
\maketitle
\markboth{Jaaniste \etal: The Supercluster--Void Network V}{}

\begin{abstract} 

  We present a study of the shape, size, and spatial orientation of
  superclusters of galaxies.  Approximating superclusters by triaxial
  ellipsoids we show that superclusters are flattened, triaxial
  objects.  We find that there are no spherical superclusters. 
  The sizes of superclusters grow with their richness:
  the median semi-major
  axis of rich and poor superclusters (having $\ge$8 and $<8$
  member clusters) is 42 and 31~\Mpc\ , respectively. Similarly, the
  median semi-minor axis is 12 and 5~\Mpc\ for rich and
  poor superclusters. The spatial orientation of superclusters, as
  determined from the axes of ellipsoids, is nearly random.  We do not
  detect any preferable orientation of superclusters, neither with
  respect to the line of sight, nor relative to some other outstanding
  feature in the large scale structure, 
  nor with respect to the directions of principal axes of
  adjacent superclusters.


\end{abstract}
\maketitle
\section {Introduction}

Superclusters of galaxies represent the largest relatively isolated
density enhancements in the Universe with a characteristic size up to
$\approx 100$ \Mpc\ ~ (Oort 1983, Einasto \etal 1994, hereafter EETDA).
Superclusters evolve from density perturbations near the
maximum of the power spectrum (Frisch \etal 1995). Thus the study of
superclusters, their shapes, dimensions, spatial distribution and
orientations gives us information about the formation and evolution of
the Universe on the largest scales.

In the present series of papers we investigate the properties and the 
distribution of superclusters traced by rich clusters of galaxies. In
the first paper of the series (Einasto \etal 1997b, Paper I) we
constructed a new catalogue of superclusters up to a redshift of $z =
0.12$; this catalogue contains 220 superclusters with at least two
member clusters. In Paper I we also investigated the large-scale
distribution of superclusters, and showed that superclusters and voids
form a rather regular network with a characteristic scale of about
$120$ \Mpc.  In the following papers we determined the correlation
function of clusters of galaxies (Einasto \etal 1997c,d, Papers II and
III), and the power spectrum of clusters of galaxies (Einasto \etal
1997a).

In the present paper we concentrate on the properties of superclusters
themselves: their form, dimension and orientation.  So far the shapes
and orientations of superclusters have been studied only in a few
papers. West (1989) analysed 11 superclusters with at least 5
members, Plionis \etal (1992) investigated the shapes of a small
number of superclusters, and Zucca \etal (1993, \ha ZZSV) studied
line-of-sight orientations of superclusters, as well as velocity
dispersions of clusters in superclusters, but not the supercluster shapes.

Superclusters can be approximated by triaxial ellipsoids, and we shall
determine both the length and the orientation of major and minor
semi-axes of these ellipsoids.  
  In Papers II and III of this series we describe how
  the correlation length of our cluster samples depend on 
  the richness of superclusters they belong to. 
In this paper we analyse the sizes of superclusters of different richness.

We also study the large-scale orientations of superclusters. 
  We shall check for the presence of a preferred orientation of
  superclusters with respect to the line of sight. 
Such an orientation could indicate the presence of contaminations
in the catalogue of clusters, or the influence of cluster peculiar
velocities and cluster movements from void interiors toward void
walls. We also look for other effects like e.g. possible correlations
in the supercluster orientation connected with well--known features of
the large--scale structure, such as the Shapley supercluster, the Bootes
void, or the Dominant Supercluster Plane described in Paper I.

So far several different kinds of preferred orientations of galaxies
and systems of galaxies have been found -- the alignment of major axes
of brightest galaxies in clusters and superclusters with the major
axis of galaxy systems and the alignment of major axes of clusters in
superclusters to be parallel with the main ridge of superclusters
(J\~oeveer and Einasto 1978, Einasto \etal 1980, Gregory \etal 1981,
Binggeli 1982, Djorgovski 1983, Flin and Godlowski 1986, West 1989).
We check for the presence of pairwise correlation for superclusters in
order to see whether such alignments extend over larger scales than
detected earlier.

The paper is organised as follows. In Section 2 we describe the
catalogue of superclusters used in this study. In Section 3 we
describe the method to find the shapes of superclusters, in Section 4
we study the properties of the superclusters (shape and size), and in
Section 5 we discuss the orientation of superclusters.  The paper concludes with
a summary of principal results.

We denote with $h$ the Hubble constant in units of 100 km s$^{-1}$
Mpc$^{-1}$.

\section {Data}

\subsection{The supercluster catalogue}

We shall use the catalogue of superclusters determined in Paper I
which is based on the compilation of all available redshift
determinations of rich clusters of galaxies, catalogued by Abell
(1958), and Abell, Corwin and Olowin (1989).  The cluster dataset has
been described by Andernach, Tago and Stengler-Larrea (1995) and
Andernach and Tago (1998). 
To construct the supercluster catalogue 
we used 1304 clusters of all richness classes up to a redshift of $z = 0.12$.
Of these clusters about 2/3 have measured redshifts, while for the rest
photometric distances have been estimated on the basis of the 10th
brightest member of the cluster.  The catalogue of superclusters in
Paper I contains all systems with at least 2 member clusters, in total
220 superclusters.  For details of the whole catalogue we refer to 
Paper I.  For the present study we use only superclusters with at
least 5 members, and with a minimum of three members having
spectroscopically measured redshifts, in total 42 superclusters.  In
Table A1 we give basic data for the superclusters analysed below: the
identification number and multiplicity (as given in Paper I), and data
on the shape and orientation.

In Paper I we showed that rich superclusters are located 
in chains and walls, the
most prominent of these being the Dominant Supercluster Plane. This
plane is almost parallel to the supergalactic YZ plane and lies in a
narrow supergalactic X interval, $ -75 < X < 25$~\Mpc. In Section 5 we
study the orientations of superclusters with respect to this plane.

\section {Ellipsoid of concentration}

\subsection {Method}

Here we describe the method we applied to determine the shape and the
size of rich superclusters, and its stability.

Superclusters are not regular systems with well-defined boundaries but
aggregates of quite sparsely distributed clusters with some central
concentration.  Therefore we have to find a mathematical method to
determine a geometrical figure or body that approximates the
distribution of clusters in superclusters.  One of such bodies, a
rectangular brick, aligned in the direction of supergalactic
coordinates, was used in \ee~ to estimate the size of superclusters.
However, this approach does not give us any information about the
orientation of superclusters.

For our purpose the 3-dimensional ellipsoid is more suitable.  There
are several methods to determine such ellipsoids, two of them have
been used earlier to study the symmetry of clustering.  Jaaniste
(1982) used the ellipsoid of concentration to study the distribution of the
companions of our Galaxy; West (1989) used the inertia ellipsoid to
study the morphology of superclusters.

In the present study we use the classical  ellipsoid of concentration
(see e.g. Korn and Korn  1961), centred on the geometrical centre
of the supercluster:
$$
\sum_{i,j=1}^3\left(\lambda_{ij}\right)^{-1}x_ix_j=5,  \eqno(1)
$$
where
$$
\lambda_{ij}={1\over{N_{cl}}}\sum_{l=1}^{N_{cl}} 
{(x^l_i-\xi_i)(x^l_j-\xi_j)}, \eqno(2)
$$
is the covariance matrix for the non-weighted distribution;
$\xi_i={1\over N_{cl}}\sum_{l=1}^{N_{cl}}x^l_i$ are the Cartesian
coordinates of the geometrical centre of the 
supercluster, and $N_{cl}$ is the multiplicity of the
supercluster. The formula determines a 3-dimensional ellipsoidal
surface  with the distance from the centre of the ellipsoid equal to
the rms deviation of supercluster members in the corresponding
direction. More correctly a mathematician defines the ellipsoid of
concentration so that "a uniform distribution function within 
it has the same covariance matrix as the given
distribution".  One of these ellipsoids, calculated for
the Shapley supercluster, is shown as an example in Figure~1.

\begin{figure*}[htbp] 
\epsfxsize=15cm
\hbox{\epsfbox[10 10 450 450]{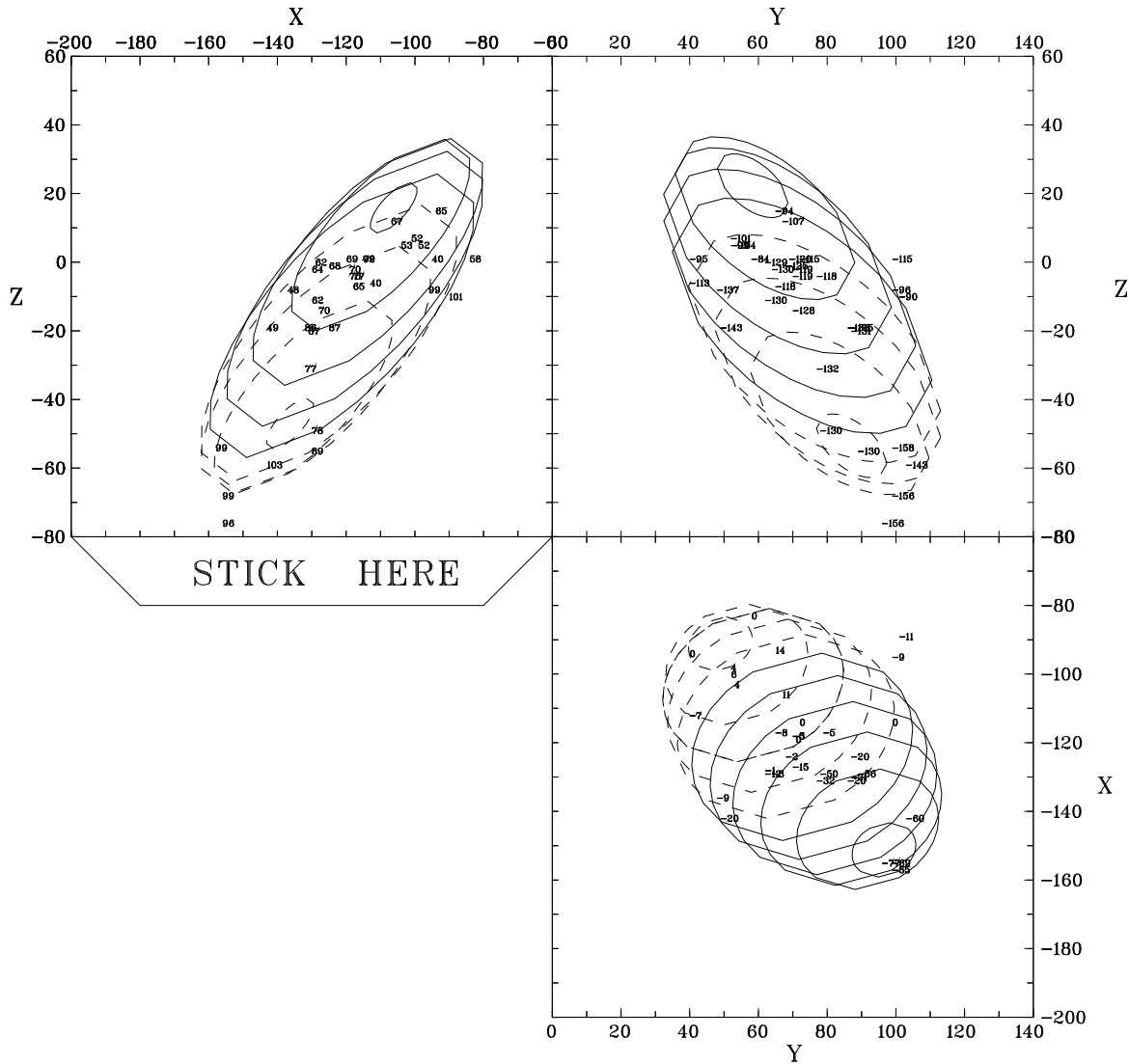}}
\caption {The Shapley supercluster ellipsoid. Ellipses are the
intersections of the ellipsoid with the planes parallel to
corresponding coordinate plane; the dashed ones correspond to
the far side of the ellipsoid. The numbers in each panel (view)
indicate the third coordinate and are plotted at the position of
the corresponding supercluster member. Six clusters seen in the
lower parts of each panel probably do not belong to the main body
of the supercluster. }
\end{figure*}

The ellipsoid, given by the standard formula of a quadric surface,
allows us to find the principal axes of the superclusters, their
spatial orientation, the spatial extent and volume for all
superclusters with $N_{cl}\ge 5$. For smaller groups the method cannot
be applied due to the small number of independent coordinates.

In Table~A1 we present the main parameters of the ellipsoid of concentration
of each supercluster studied: the reference number, $No$, from Table A1 of
Paper I; the number of member clusters, $N_{cl}$; the length of
the three principal semi-axes: $c$, $b$ and $a$; the ellipticity, $e$,
defined as the ratio of the volume of the ellipsoid to the volume of
the sphere with radius equal to the major semi-axis, $V={{4\pi}\over 3}abc$ to
$V_{sph}={{4\pi}\over 3} a^3$); the angles of the ellipsoids'
minor and major axes ($\angle cx$, and $~\angle ax $) with respect to the
supergalactic $X$-axis, and 
to the line of sight ($\angle cr$,$~\angle ar$).

\begin{figure}[htbp] 
\epsfxsize=8.cm
\hbox{\epsfbox[0 20 230 300]{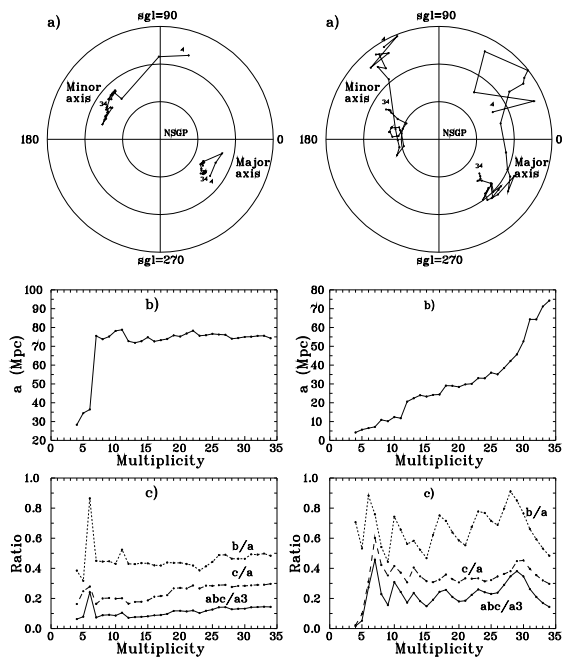}}
\caption {The variation of supercluster parameters during exclusion
of member clusters, shown here for the Shapley supercluster.  Left
   panels correspond to random exclusion, right panels to the
   exclusion of the most distant cluster of the previous
   configuration.  Panel (a) shows the direction of the major and the
   minor axis projected on the supergalactic northern hemisphere;
   panel (b) -- the size of supercluster (semi-major axis); and panel
   (c) the axial ratios and ellipticities. The dots in panel (a) mark
   the direction of the minor and major axis for the Shapley
   supercluster as a function of the number of member clusters from 34
   down to 4.
}
\end{figure}

\subsection {Stability of the method}

It is clear that parameters of the ellipsoid depend on the number and
location of clusters in the supercluster. To study this dependence we
exclude members of a supercluster one by one and recalculate the
parameters of the ellipsoid once again at each step. In this manner we
reduce the multiplicity down to a minimum $N_{cl}$ of four.  For
smaller $N_{cl}$ the ellipsoid cannot be determined. 
We studied four real and two artifical superclusters of 12 to 150
members in this way (SCL 160, 114, 9 and 124 from the catalogue, as well as
two random distributions of given symmetry with $N_{cl}=50$ and 150); 
several different realizations were done for each random exclusion.
The results are
presented for an example of the Shapley supercluster in Figure 2.

The left panels of Figure~2 show the variation of the parameters of
the ellipsoid of concentration during the process of {\it randomly}
excluding supercluster members.  We see that the ellipsoid parameters
are surprisingly stable: a reduction of the number of members 
down to 
$\approx 30$\% changes the main parameters by no more than $\pm 10$\% and
the ellipsoid axes do not drift by more than $10^\circ$ .

In the right panels of Fig.~2 the member clusters are excluded in the
order of their distance from the supercluster centre, starting with
the outermost one.  In this case we see a rapid contraction of the
ellipsoid accompanied by strong variations of its shape. However, the
directions of principal axes is quite stable until the number of
members is reduced down to about 8. The covariance matrix, and
consequently, parameters of the ellipsoid of consentration, depend strongly
on the most distant member clusters. A similar exercise of removing
the innermost members shows that in this case the main parameters 
of the ellipsoid remain nearly unchanged.

These calculations show that the main parameters of the 
ellipsoid of concentration depend primarily on the 
location of the distant members of
superclusters.  The method discussed above can be used for
superclusters with $N_{cl}\ge 8$ members.  For superclusters with
$N_{cl}=5\dots 7$ members we must take into account the fact that
random deviations of supercluster parameters due to poor determination
of membership may be quite large.

We analysed also the possible influence of observational errors on the
parameters of the ellipsoid of concentration.  Redshift errors are typically
less than $100$ km/s, and have little influence on the 
ellipsoid. Errors due to the velocity dispersion of galaxies in
clusters are larger, the median value is about $700$~km/s (Mazure
\etal 1996, and Andernach \& Tago 1998). These errors may be
important only for clusters whose mean velocity is based on 
one or a few galaxy redshifts. In those cases where the redshift is
determined from the brightest members of clusters that lie in the
dynamical centre of a cluster, these errors are much
smaller. A possible source of errors are the peculiar velocities of
clusters (Watkins 1997) and bulk motion of clusters with respect to
the Hubble flow. 
If the bulk motion exists, then we can assume that clusters in
the same supercluster have similar velocities, and that this error does not
influence our results.  To analyse the influence of peculiar velocities we
shifted supercluster members randomly in velocity space up to 
$\pm 700$~km/s, and
recalculated the supercluster parameters from the revised membership of 
individual clusters. 
We found that these factors are
noticeable only when they cause a removal of distant members
from superclusters; in this case the changes are similar to those
described above.

\subsection {Comparison with previous studies}

We compared our method with those used by West (1989) and ZZSV to
derive the shapes of superclusters.  These methods are also based on
the covariance matrix, and describe the geometrical properties of
superclusters in the same way. The main difference is the
normalisation used by West: to avoid the possible bias by the outer
members he divided local coordinates of clusters by the distance from
the centre of the supercluster. In this way he projected all members
of the supercluster to the sphere of unit radius surrounding the
geometrical centre of the supercluster. 
We were afraid that using this method all information about the spatial
dimensions of the supercluster would be lost. Nevertheless,
our calculation of supercluster shape parameters according to West's (1989)
method showed that the orientation of superclusters (i.e. the direction
of the major axes) were in most cases {\bf exactly the same} 
as those according to our method ; 
we found only three superclusters (of 42)
where two of the main axes changed their order.

Differences in shape between the two methods are somewhat greater 
but not crucial: the
ellipticities $V/V_{sph}=bc/a^2$ differ by only a factor of 1.5  on
average (superclusters are more elliptical according to our method). 
We conclude that the advantage of our method 
compared to West's method is the
estimation of spatial dimensions of superclusters.

\section {Geometrical properties of superclusters}

In this Section we analyse geometrical properties of superclusters --
their size and shape. For this purpose we use the length of the semi-major
axis, $a$, the intermediate semi-axis, $b$, the semi-minor axis , $c$, and
the ratios of axes, $b/a$ and $c/a$ and the ellipticity, $e$. 

\begin{figure}[htbp] 
\epsfxsize=6.cm
\hbox{\epsfbox[0 20 240 240]{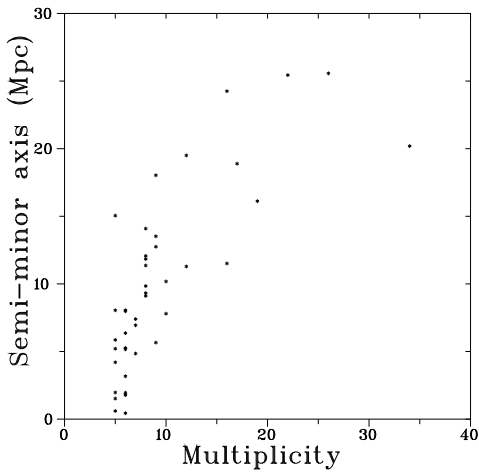}}
\caption {The length of the semi-minor axis of the
superclusters of different multiplicity $N_{cl}$.}
\end{figure}

Mean sizes of semiaxes for superclusters with multiplicity $N_{cl} <
8$ and $N_{cl} \geq 8$ are given in Table~1.  In Figure~3 we show
the semi-minor axis of the superclusters as a function of their richness.

The median length of the semi-minor axes $c$ is 10 \Mpc, close to
the value found by \ee.  The semi-minor axis of the ``thinnest''
superclusters (i.e. those with smallest value of $c$ or with the
highest ellipticity) is only 0.4 \Mpc, of the order of the radius
of a normal Abell cluster. This supercluster, SCL 126, consists of
a single elongated cluster filament; we shall discuss this
supercluster in more detail in Section 5.  The median value of the
intermediate semiaxis $b$ is 20 \Mpc.

\begin{table}
\caption{Median semiaxes (a and c) of superclusters and the correlation
lengths ($r_0$) for clusters in superclusters (from paper II)}
\begin{tabular}{rlllllll}
\hline
Multiplicity& $a$ & $c$ & $r_0$ \\  
              &\Mpc\  & \Mpc\  & \Mpc\ \\
\hline
$N_{cl} < 8$     &  31  &   5  & 17 \\
$N_{cl} \geq 8$  &  42  &  12  & 46 \\
\hline
\end{tabular}
\end{table}

Table 1
shows that the supercluster sizes depend strongly on their richness.
A similar tendency was found for the correlation length, $r_0$ (Papers
II and III; the correlation length from Paper II is also given in the
Table). In superclusters of small and medium richness the correlation
length is approximately equal to the geometric mean of the median
values of the major and minor semiaxes while in very rich superclusters 
it is near to
the median value of the semi-major axis. Thus we see that the
correlation length characterises the size of superclusters in a slightly
different manner than the size of semiaxes. 

The location of superclusters in the $c/a$ vs $b/a$ plane is shown
in Figure~4. We see that there are no superclusters with spherical
shape.  One of the roundest superclusters have axial ratios a:b:c = 1:0.5:0.5 
(the supercluster in Microscopium with ten members, SCL 174); the
maximum density of points in Figure~4 corresponds to regions of
axes ratios between 1:0.5:0.2 and 1:0.6:0.4.  The median ellipticity is
0.125; 40\% of superclusters have $e<0.1$, 75\% less than 0.2, and
there are no superclusters with $e>0.4$.

Based on data for 11 superclusters West (1989) found that
superclusters are flattened with typical axial ratios of 1:0.3:0.3 to
1:1:0.3 which agrees with  our results rather well.

\begin{figure}[htbp] 
\epsfxsize=7.cm
\hbox{\epsfbox[-10 30 200 220]{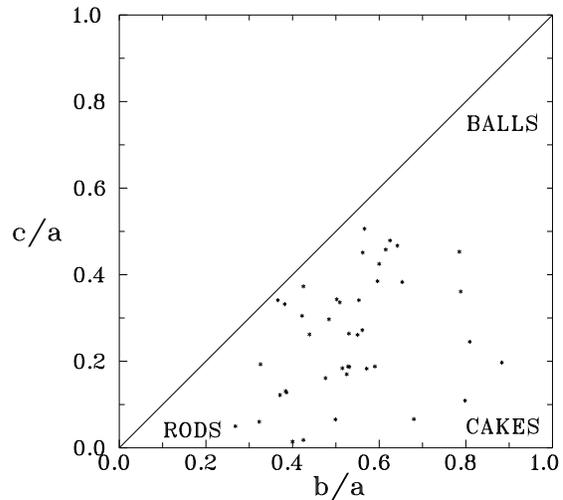}}
\caption {The supercluster axis ratios  $c/a$
  versus $b/a$. The corners of the triangle correspond to 1:0:0  
  (prolate shape), 1:1:1 (spherical shape) and 1:1:0 (oblate shape).}
\end{figure}

Thus our results show that the superclusters are quite flattened
objects: the shape of superclusters can be described as ``prolate'' or
``rods'', and ``oblate'' or ``cakes''. Elongated ``oblate''
superclusters are more common than flattened ``prolate'' ones.

\section {Large scale orientation of superclusters}

The strong elongation of superclusters allows us to find the spatial
orientation of superclusters, and to study the correlation of the
orientation with respect to some preferred direction, like the location of
the observer, adjacent rich superclusters or voids, etc.

As pointed out above, we have no spherical superclusters in our
sample. Consequently, the orientation of superclusters can be
determined by the directions of the major and minor axes.  We shall
calculate the distribution of the angles of supercluster axes with
respect to directions mentioned above, and compare this with a
uniform (or random) distribution.

\subsection {Line-of-sight orientation}

  First we shall test for the presence of a possible observational bias 
that could reflect artefacts 
  in the cluster catalogue, for instance the presence of
  projection effects in the cluster catalogue, or the influence of cluster
  peculiar velocities in superclusters.  
Distances of clusters are
determined from their redshifts. Thus, if clusters in superclusters
have a large dispersion of velocities with respect to the supercluster
centres, then, using our method, superclusters will be elongated 
along the line of sight.
Such a ``finger of God'' effect is observed in clusters of
galaxies. Clusters are virialised objects and have large internal
velocity dispersion of galaxies.  Due to the large sizes of
superclusters we do not expect the latter effect to be strong.

In Figure~5 we present the line-of-sight orientation of all
superclusters with $N_{cl} \geq 5$. We see that both major and minor
axes have a nearly isotropic distribution. The only deviation from the
uniform distribution is a small excess at small values of cosines
(histogram above the diagonal) corresponding to the major axis lying
nearly perpendicular to the line of sight ($\cos (ar)<0.3$,
i.e. $\angle ar>70^\circ$). Surprisingly, among these ``nearly
perpendicular'' superclusters there are three of the four most
flattened systems (SCL 18, 126 and 202) -- a fact, suggesting a
possible influence of large--scale motions on the morphology of
superclusters in redshift space. We discuss this phenomenon in Section 6. 
In any case no preferred line-of-sight orientations are present in our
sample.  The same was found also by West (1989) and by Zucca \etal
(1993).

\begin{figure}[htbp] 
\epsfxsize=10.cm
\hbox{\epsfbox[-10 20 350 450]{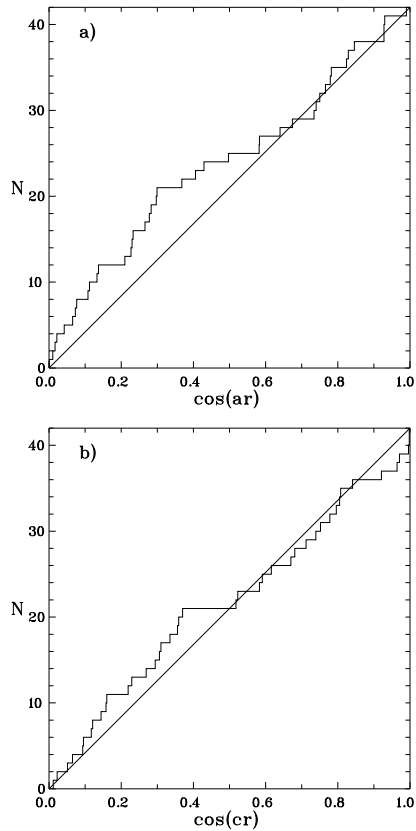}}
\caption {The integrated distributions of cosines of
angles of major axes (upper panel) and minor axes (lower panel) with
respect to the line-of-sight vector.  Small values of cosines
(histogram above the diagonal) corresponds to the tendency for axes to
be perpendicular to the main direction (line-of-sight vector); high
values (histogram below the diagonal) suggest the directions to be
parallel to the line of sight.}
\end{figure}

\subsection {Orientation with respect to voids and superclusters}

It is well known from observations and numerical simulations that the
matter flows away from voids towards centres of large mass
concentrations (Kaiser 1987, Gramann \etal 1993).  Superclusters are
the largest known mass concentrations in the Universe, thus one may
assume that they are oriented with respect to adjacent voids (centres
of divergence), or nearby superclusters (centres of convergence).  For
this purpose we checked the orientation of superclusters relative to
the three richest superclusters in our catalogue with
$N_{cl}>20$. These very rich superclusters are the Shapley
supercluster, the Sculptor supercluster, and the Horologium
supercluster.  To investigate the orientation of superclusters
relative to voids we used the catalogue of voids by EETDA.

The results of this analysis showed no significant deviation from
isotropy.  The only deviation from the isotropic distribution is a
weak tendency of perpendicularity of major axes to the direction
toward the centre of the Bootes void.

The absence of preferable orientations with respect to voids is somewhat
surprising. Superclusters surround voids, thus at least superclusters
at the edge of voids should be perpendicular to the direction of the
void centres.  However, a supercluster oriented perpendicular with
respect to one void may be parallel to another one causing together
the absence of preferred orientations.  Our negative result may also
be due to the small number of superclusters: there are 42
superclusters and 26 voids in our catalogue.

\subsection {Global trends in orientation}

Global orientations have been searched for in two ways: studying the
distribution of direction cosines in Cartesian space, and searching
the possible centre of convergence of the supercluster axes.  We
select the supergalactic $X-$axis as the main direction because it is
perpendicular to the Dominant Supercluster Plane (DSP, see Paper
I). However, we found no preferential orientation, except for a very
weak tendency of minor axes to be parallel to the $X-$axis
(perpendicular to the DSP).

We tried to find a possible ``centre of convergence'', independent of
the visible structures in our sample.  Such a centre could indicate
the presence of a very large mass concentration outside our sample
boundaries.  We searched for such a centre in the following way: first
we calculated the direction cosines of principal axes for all
supercluster ellipsoids; then we derived the meeting-points (points
at which the distance between the two lines is the shortest) of major
axes for all possible pairs of superclusters. Thus we obtained 861
meeting-points for 42 superclusters and tried to find the common
meeting-point as a concentration of them using cluster analysis. We
found this point to lie close to the coordinate center, i.e. at the
location of the observer; so the distribution of line-of-sight angles
from this ``new'' viewpoint is almost the same as in Figure~5. We
conclude that no such point of convergence exists; the directions of
axes of the superclusters are distributed isotropically, and the
clustering of meeting--points only reflects the central symmetry of
our sample.

\subsection {Orientation of supercluster pairs in supercluster-void
network}

As we showed in Paper I, superclusters form a rather regular
network with characteristic distance of about $120$~\Mpc. This
distance corresponds to the distance  between superclusters at
opposite void walls; about 75\% of very rich superclusters 
have a neighbouring supercluster at this distance (typically, the
second nearest neighbour).

As we mentioned in the Introduction, several different kinds of
preferred orientations of galaxies and systems of galaxies have
been found, among them the alignment of major axes of clusters in
superclusters with the main ridge of superclusters. 
Therefore we checked whether such alignments extend over
scales that characterize the supercluster-void network. To this aim we
computed the orientations of superclusters with respect to their
nearest three neighbours.

\begin{figure}[htbp] 
\epsfxsize=18.cm
\hbox{\epsfbox[10 30 550 650]{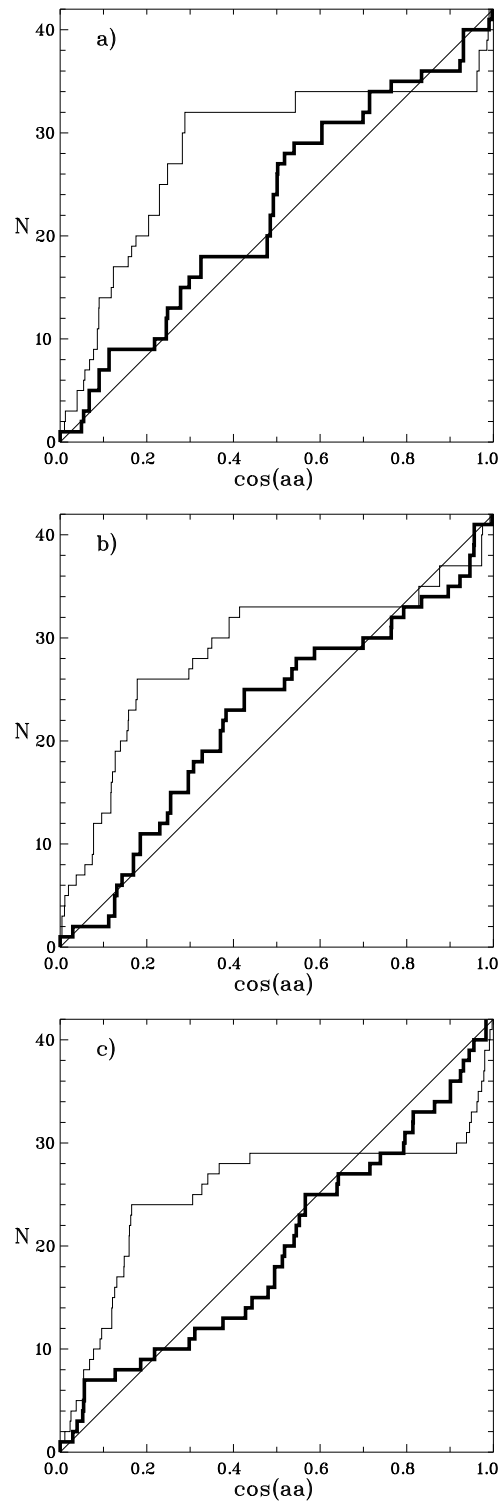}}
\caption {Orientation of supercluster pairs: 
cosine of angles between major axes of neighbouring superclusters for
the nearest neighbour (panel a), the second nearest (panel b) and the third
nearest (panel c) neighbour.  The thick histogram corresponds to
observations, the thin histogram to model superclusters  (see
text), and the thin straight line shows a uniform distribution. }
\end{figure}

To test the method of finding supercluster orientations, as well as
for comparison with orientations of real superclusters, we
additionally generated a set of superclusters located on a regular
3-dimensional lattice oriented in the direction of
the coordinate axes. In these sets the number of superclusters was
equal to the number of observed superclusters, 42. The multiplicity
matched the observed one, and the shapes of superclusters were
elongated. The major axes of superclusters were aligned along the
rods of the lattice (in the direction of three coordinate axes);
the exact location of superclusters along the rods was random (we
only avoided overlapping of superclusters located in the same
direction on the same rod).  Therefore, for about 2/3 of
superclusters the nearest neighbours are perpendicular to each
other, and for the rest of them the nearest neighbours are
parallel. Then we calculated orientations between real supercluster
pairs, as well as between model superclusters (Figure~6). 

As is seen in Figure~6, the relative orientations between model
superclusters elongated along rods of a regular lattice show strong
perpendicular and parallel orientations. Thus our method is able to
detect preferable orientations. As shows the Kolmogorov-Smirnov
test, the possibility that model superclusters could have random
orientations can be rejected at 99\% confidence level. 
Figure~6 shows also that the orientations between real supercluster
pairs in the supercluster-void network are nearly random. The same is
confirmed by the Kolmogorov-Smirnov test.

Summarising, we do not find any correlation or regularity in the
spatial orientation of superclusters. The orientation of superclusters
seems to be isotropic on all scales up to the sample volume.

\section {Discussion} 

The fact that we have found no sign of peculiar motions of clusters
within rich superclusters is not surprising since expected velocities
must be of the order of the observationally known bulk motions of
galaxies, 600~km/s (Branchini and Plionis 1996). 
An object having a peculiar velocity of 600~km/s, will travel about
6~\Mpc\ in a Hubble time relative to the general Hubble flow.
Dimensions of superclusters exceed these distances many times.
Numerical experiments also indicate that superclusters have the form
and shape which is close to the form and shape during their formation
(Frisch \etal 1995).

The isotropy of supercluster orientation with respect to the line of
sight confirms that superclusters are still in the stage of formation.
The lack of a line-of-sight elongation excludes the idea of
gravitationally bound relaxed superclusters, where internal motions
would lead to radially elongated ``fingers of God''. 

The lack of spherical superclusters is in harmony with the view that
superclusters are not dynamically bound systems but parts of the
supercluster-void network which as a whole is expanding with the
Hubble flow.  Numerical experiments show that in high-density regions
clusters and groups of galaxies are concentrated to long chains. This
phenomenon was detected empirically many years ago (J\~oeveer and
Einasto 1978, J\~oeveer \etal 1978).

A weak tendency of major axes to be perpendicular to the line of sight
is in agreement with theoretical predictions.  
Supercluster members are moving towards the centre of the
supercluster, and line-of-sight dimensions of superclusters are
decreased in redshift space (Kaiser 1987, Gramann \etal 1993).

In this connection we note that three very flat superclusters,
mentioned in Section 3, are very elongated systems of 5 or 6
members. These are SCL 18, a relatively nearby supercluster in Phoenix
($d \approx 82$~ \Mpc), and superclusters 126 and 202, both at a
distance of about 240~ \Mpc. The list of members of these
superclusters is given in Paper I.  The dimensions of these
superclusters in the plane of the sky exceed their line-of-sight
dimension between 10 times (for SCL~18) and 30 times (for SCL~126). 
We note that all distances of their member clusters are
determined spectroscopically with great accuracy.  We suppose that the
perpendicularity of these superclusters to the line of sight is enhanced
and we see the effect of the matter inflow towards the centers of the
superclusters.  The possibility of such an effect has been discussed,
for example, in Praton \etal (1997), where they show that the redshift
effects may selectively enhance the appearance of the superclusters
lying perpendicular to the line of sight. This shape reflects also the
formation history of superclusters in the sense that clusters may be
located along filaments and sheets (Colberg \etal 1997).  A detailed
dynamical study of these regions
appears worth-while: as the nearest of them is at a distance of 82~\Mpc, 
the concentration of non-cluster galaxies
to this plane should be visible.

\section {Conclusions}

We have studied geometrical properties of superclusters approximating
their shape with an ellipsoid of concentration.  Our main results can be
summarised as follows.

i) Superclusters are elongated systems with characteristic axial
ratios between 1:0.5:0.2 and 1:0.6:0.4. There are no spherical superclusters.

ii) We found no significant orientation of superclusters, neither towards the
observer, nor the large voids, nor towards any other large-scale feature.
Superclusters are oriented isotropically in space.  There are no strong
contaminations in the cluster catalogue that could cause preferable
orientations of superclusters along the line of sight.

iii) The size of superclusters depends on their richness. This effect
is similar to the dependence of the correlation length of clusters on
the richness of superclusters they are located.

\acknowledgements

We thank the referee, Dr. G. Zamorani, for many useful comments
which helped improve the paper.
This work was supported by the Estonian Science Foundation (grant
182), by the International Science Foundation (grant LLF100), and by
the Max-Planck-Gesellschaft. JJ would like to thank the Division of
Astronomy and Physics of Estonian Academy of Sciences for financial
support; JE was supported in Potsdam by the Deutsche
Forschungsgemeinschaft; HA benefitted from financial support by
CONACYT (Mexico; C\'atedra Patrimonial, ref 950093).

\vfill\eject

\onecolumn
\appendix{Appendix}

{\bf Table~A1.} The shape and orientation of superclusters
$$\vbox
{\small
{\tabskip=7pt plus3pt minus4pt 
\halign to 17truecm { \hfil# & \hfil# & \hfil#
& \hfil# &
 \hfil#&\hfil# &\hfil# &\hfil#&\hfil# &\hfil# &\hfil# &\hfil# 
\cr
\noalign {\smallskip}
\noalign{\hrule width 17truecm height 0.4pt}
\noalign{\smallskip}
(1)& (2)~~ & (3) & (4)~~ &(5)~~ & (6)~~ & (7) & (8) & (9) & (10)~~
\cr
\noalign{\smallskip}
 No&$N_{CL}$&  
$c~~~$&$b~~~$&$a~~~$&$V/V_{sph}$& 
$\angle (cx)$
&$\angle (ax)$  
&$\angle (cr)$&$\angle (ar)$ \cr 
  SCL &  &
~~Mpc~&~Mpc~~&~~Mpc~~&&deg &deg &deg&deg
 \cr
\noalign{\smallskip}
\noalign{\hrule width 17truecm height 0.4pt}
\noalign {\smallskip}
3 &   8
 &   9.1 &  25.8 &
  48.6 &  .099 &
  60 &  77 &
  72 &  41 \cr
5 &   5
 &   2.0 &  15.0 &
  30.1 &  .033 &
  26 &  65 &
  84 &  76 \cr
9  &  22
 &  25.4 &  29.2 &
  76.6 &  .127 &
  87 &  42 &
  41 &  68 \cr
10 &  17
 &  18.9 &  20.3 &
  55.5 &  .125 &
  56 &  45 &
  37 &  54 \cr
18 &   6
 &   1.8 &  18.2 &
  26.7 &  .045 &
  75 &  25 &
  77 &  86 \cr
24 &   7
 &   6.9 &  21.8 &
  36.9 &  .111 &
  63 &  41 &
  47 &  83 \cr
30 &   8
 &  14.1 &  21.4 &
  42.0 &  .171 &
  44 &  59 &
  80 &  73 \cr
34 &   6
 &   5.3 &  15.5 &
  32.6 &  .077 &
  89 &  29 &
  76 &  88 \cr
48 &  26
 &  25.6 &  35.1 &
  54.7 &  .300 &
  88 &  68 &
  74 &  42 \cr
49 &   5
 &   5.2 &  16.2 &
  28.5 &  .105 &
  83 &  25 &
   3 &  87 \cr
65 &   5
 &  15.1 &  19.6 &
  31.5 &  .299 &
  37 &  52 &
  88 &  66 \cr
67 &   6
 &   5.2 &  15.1 &
  39.4 &  .050 &
  38 &  57 &
  51 &  38 \cr
91  &   9
 &  13.5 &  21.9 &
  39.6 &  .189 &
  60 &  64 &
  54 &  77 \cr
93 &   9
 &   5.7 &  17.2 &
  46.3 &  .045 &
  60 &  43 &
  38 &  85 \cr
107 &   8
 &  12.1 &  17.0 &
  28.4 &  .255 &
  22 &  70 &
  87 &  21 \cr
109 &   8
 &  11.8 &  17.3 &
  34.6 &  .172 &
  31 &  68 &
  83 &   8 \cr
111 &  16
 &  24.2 &  42.1 &
  53.6 &  .355 &
  77 &  79 &
  13 &  76 \cr
114 &  16
 &  11.5 &  38.1 &
  47.1 &  .198 &
  86 &  17 &
  42 &  47 \cr
124 &  34
 &  20.2 &  32.9 &
  67.9 &  .144 &
  49 &  59 &
  47 &  42 \cr
126 &   6
 &    .4 &  12.5 &
  31.3 &  .006 &
  56 &  83 &
   4 &  89 \cr
127 &   5
 &   4.2 &  18.8 &
  21.2 &  .174 &
  29 &  76 &
  15 &  76 \cr
128 &   6
 &   2.0 &  10.5 &
  32.4 &  .019 &
  46 &  44 &
  69 &  40 \cr
138 &  12
 &  19.5 &  26.2 &
  42.6 &  .282 &
  47 &  81 &
  68 &  21 \cr
150 &  10
 &  10.2 &  17.4 &
  26.6 &  .250 &
  64 &  45 &
  58 &  82 \cr
157 &   6
 &   8.0 &  11.1 &
  26.2 &  .129 &
  48 &  45 &
  70 &  85 \cr
158 &   8
 &   9.8 &  16.5 &
  37.6 &  .115 &
  68 &  63 &
  72 &  38 \cr
160 &  12
 &  11.3 &  19.0 &
  58.3 &  .063 &
  72 &  26 &
  84 &  60 \cr
164 &   5
 &   5.9 &  18.1 &
  34.5 &  .089 &
  61 &  57 &
  36 &  54 \cr
168 &   5
 &   1.5 &   8.1 &
  30.3 &  .013 &
  43 &  57 &
  80 &  21 \cr
170 &    7
 &   4.8 &  13.6 &
  26.4 &  .095 &
  54 &  40 &
  69 &  34 \cr
174 &  10
 &   7.8 &   9.7 &
  17.3 &  .253 &
  69 &  86 &
  71 &  50 \cr
188 &   9
 &  12.8 &  14.5 &
  34.1 &  .159 &
  89 &  25 &
   5 &  89 \cr
190 &   5
 &   8.0 &  17.0 &
  30.8 &  .144 &
  73 &  41 &
  81 &  73 \cr
192 &   8
 &  11.4 &  24.8 &
  31.5 &  .284 &
  84 &  67 &
  36 &  64 \cr
193 &   8
 &   9.3 &  19.2 &
  34.2 &  .153 &
  31 &  58 &
  83 &  72 \cr
197 &   9
 &  18.0 &  20.1 &
  35.6 &  .286 &
  20 &  70 &
  53 &  74 \cr
202 &   5
 &    .6 &  14.1 &
  33.3 &  .008 &
  14 &  84 &
  44 &  83 \cr
205 &  19
 &  16.1 &  32.4 &
  61.1 &  .140 &
  12 &  85 &
  86 &  32 \cr
208 &   6
 &   6.4 &  19.1 &
  49.5 &  .050 &
  54 &  37 &
  22 &  72 \cr
209 &   7
 &   7.4 &  11.5 &
  19.2 &  .230 &
  57 &  66 &
  32 &  82 \cr
210 &   6
 &   8.0 &  22.6 &
  42.9 &  .099 &
  26 &  75 &
  58 &  88 \cr
213 &   6
 &   3.2 &  23.2 &
  29.1 &  .087 &
  65 &  79 &
  89 &  34 \cr
}}}$$

\msn
Columns are as follows:
\msn
1) reference number from Table A1 in Paper I $No$;
\msn
2) number of clusters $N_{cl}$;
\msn
3 - 5) length of the three principal semiaxes of 
ellipsoid of concentration, $c$, $b$ and $a$; 
\msn
6) the ratio of the volume of the ellipsoid to the
volume of a sphere of the same diameter $V/V_{sph}$;
\msn
7, 8) angles of axes with the
supergalactic $X$-axis $\angle cx$, $~\angle ax $; 
\msn
9, 10) angles of the minor and the major axis with the line-of-sight
direction $\angle cr$, $~\angle ar$. 

\end{document}